\documentclass[graybox]{svmult}
\usepackage{amsmath}
\usepackage{mathptmx}       % selects Times Roman as basic font
\usepackage{helvet}         % selects Helvetica as sans-serif font
\usepackage{courier}        % selects Courier as typewriter font
\usepackage{type1cm}        % activate if the above 3 fonts are
\usepackage{makeidx}         % allows index generation
\usepackage{graphicx}        % standard LaTeX graphics tool
\usepackage{multicol}        % used for the two-column index
\usepackage[bottom]{footmisc}% places footnotes at page bottom

\makeindex             % used for the subject index
\usepackage{natbib}
\def\apj{ApJ}
\def\mnras{MNRAS}
\def\aj{AJ}
\def\apjl{ApJ Lett}
\def\araa{ARAA}

\def\nat{Nature}

\begin{document}
\bibliographystyle{spbasic}
%\texttt{vecphys}

\title*{Origins and Interpretation of Tidal Debris}
\author{Kathryn V. Johnston}
\institute{Columbia University
}
\motto{Chapter~6 from the volume ``Tidal Streams in the Local Group and Beyond: Observations and Implications''; ed. Newberg, H.~J., \& Carlin, J.~L.\ 2016, Springer International Publishing, Astrophysics and Space Science Library, 420\\ ISBN 978-3-319-19335-9; DOI 10.1007/978-3-319-19336-6\_1}

\maketitle

\abstract{The stellar debris structures that have been discovered around the Milky Way and other galaxies are thought to be formed from the disruption of satellite stellar systems --- dwarf galaxies or globular clusters --- by galactic tidal fields. 
The total stellar mass in these structures is typically tiny compared to the galaxy around which they are found, and it is hence easy to dismiss them as inconsequential. However, they are remarkably useful as probes of a galaxy's history (as described in this chapter) and mass distribution (covered in a companion chapter in this volume). This power is actually a consequence of their apparent insignificance: their low contribution to the overall mass makes the physics that describes them both elegant and simple and this means that their observed properties are relatively easy to understand and interpret.
}

\section{Introduction}
\label{sec:intro}

The debris structures described in this volume that have been discovered around the Milky Way and other galaxies are thought to be formed from the disruption of satellite stellar systems --- dwarf galaxies or globular clusters --- by galactic tidal fields. 
The total stellar mass in these structures is typically tiny compared to the galaxy they might be found around and it is easy to dismiss them as visually striking, but inconsequential. However, they are remarkably useful as probes of a galaxy's history (covered in this chapter) and mass distribution (covered in the next chapter). This power is actually a consequence of their apparent insignificance: their low contribution to the overall mass makes the physics that describes them both elegant and simple and this means that their observed properties are relatively easy to understand and interpret.

\begin{figure}[!t]
%\sidecaption
%\includegraphics[scale=0.2]{johnston_figs/fig1.eps}
\includegraphics[width=1.0\textwidth]{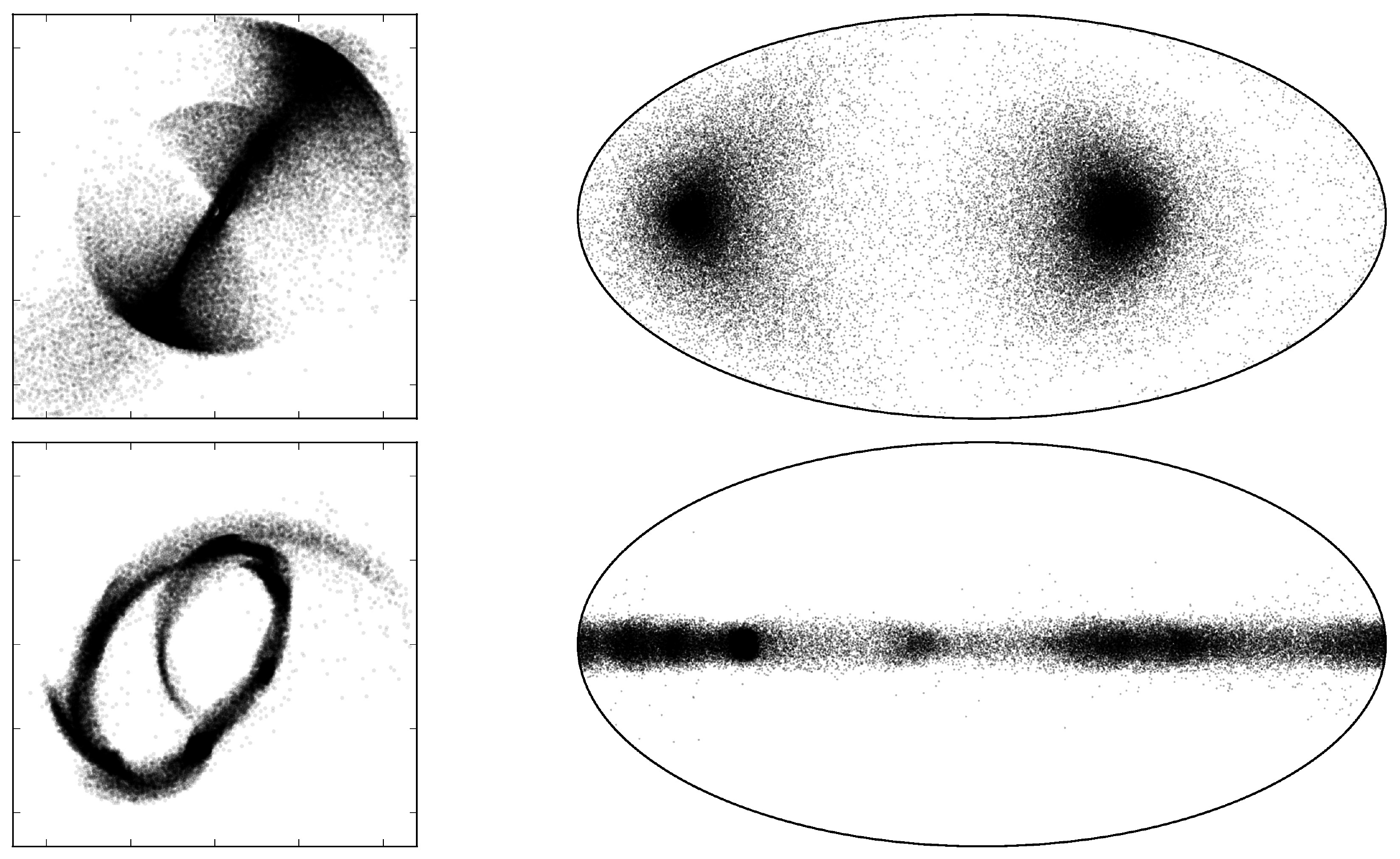}
\caption{External (left panels) and internal (i.e. all-sky projections -- right panels) views of two examples of debris structures seen in numerical simulations described in \S \ref{sec:nbody}. Both debris structures were formed along orbits with the same energy as a circular orbit at radius 25 kpc in the parent potential. The top panels are for a $6.5 \times 10^8 M_\odot$ satellite falling in along a highly eccentric orbit (angular momentum $L/L_{\rm circ}=0.1$) and the bottom panels are for the same satellite on a more mildly eccentric orbit ($L/L_{\rm circ}=0.9$).}
\label{fig:structures}   
\end{figure}

Figure \ref{fig:structures} contrasts internal and external views of two examples of debris structures seen in numerical simulations: streams\index{streams} from the disruption of a satellite along a mildly eccentric orbit (lower panels) and shells\index{shells} from the disruption of the same satellite along a much more eccentric orbit (upper panels; see \S \ref{sec:nbody} for a more detailed description of these simulations).
This chapter first outlines our understanding of the formation of such debris structures (\S \ref{sec:principles})
and then goes on to explore what we can learn about dying and long-dead satellites from observations of their debris (\S \ref{sec:observables}) and the implications of the cosmological context for properties of the combined system of all debris structures that form our stellar halo (\S \ref{sec:cosmology}).

The following chapter discusses what debris structures can tell us about the tidal field --- i.e. the mass distribution of the galaxy that destroyed them.

\section{Illustrative N-body simulations}
\label{sec:nbody}

\begin{figure}[t]
%\sidecaption
%\includegraphics[scale=0.4]{johnston_figs/fig2.eps}
\includegraphics[width=1.0\textwidth]{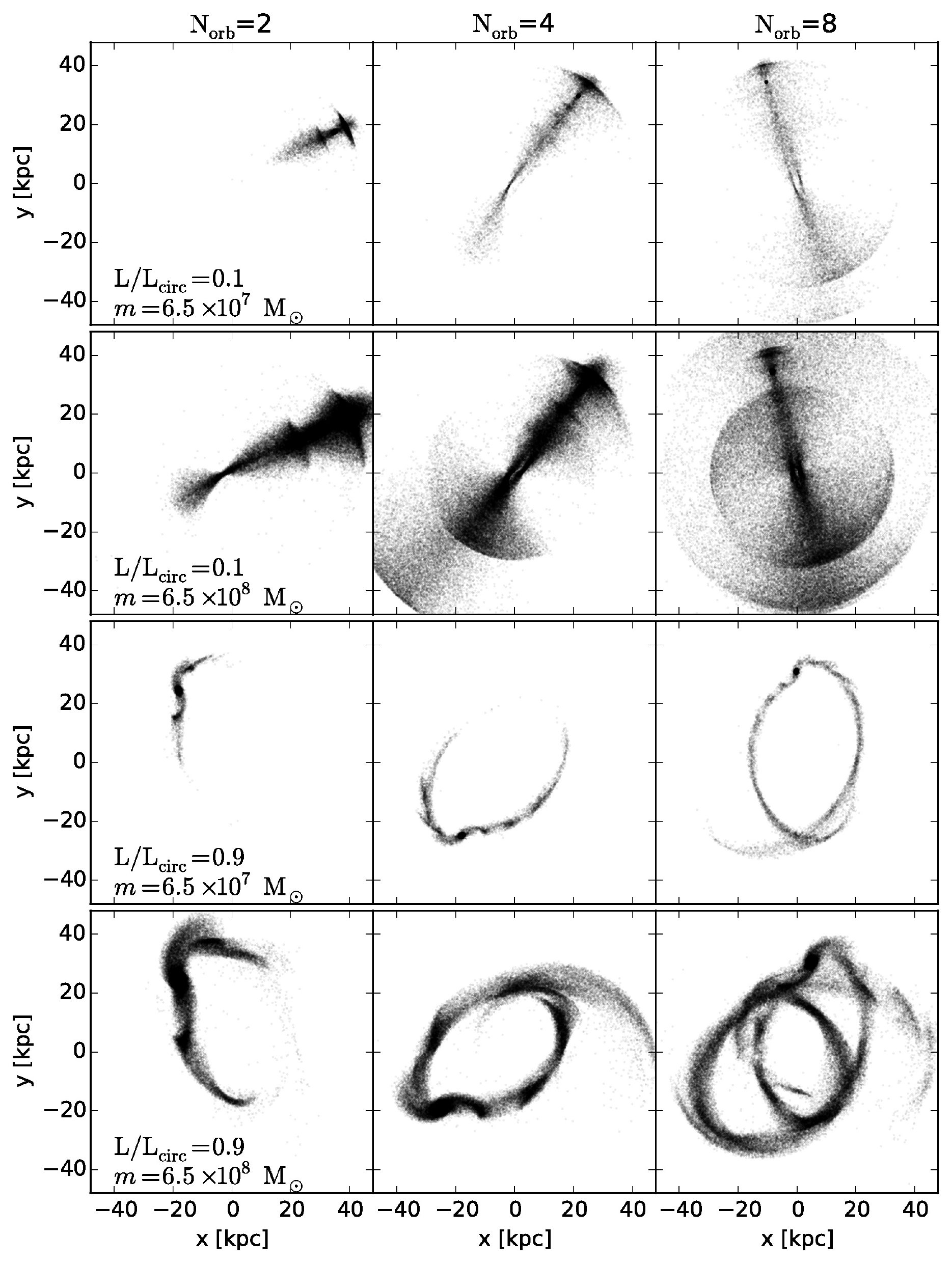}
\caption{Evolution of debris along highly (top panels) and mildly (bottom panels) eccentric orbits for two different mass satellites (labelled in left-hand panel).  Note that highly eccentric orbits produce shells, while more circular orbits form streams.}
\label{fig:evol}   
\end{figure}

The descriptions of debris evolution in this chapter are illustrated with a set of N-body simulations\index{N-body simulations} \citep[see][for a review of techniques]{dehnen11} of satellite disruption presented in \citet{hendel15}.
 The self-gravity of the satellite in these  simulations was calculated with the Self-Consistent
Field (SCF) code, which uses basis function expansions to represent the mutual influence of the particles on each other  \citep{hernquist92}.
 In each simulation, a $10^5$
particle NFW-profile\index{Navarro, Frenk, \& White (NFW) profile} \citep{nfw96} satellite was inserted at the apogalacticon of
its orbit in a static, spherical host
halo, with characteristics of dark matter halos thought to host Milky-Way-sized galaxies \citep[NFW profile with a virial mass of $M=1.77 \times 10^{12} M_\odot$ and a scale radius of 24.6 kpc, see][]{nfw96}. 
The satellite was evolved first in isolation,
then the host potential was turned on slowly over 10 satellite
internal dynamical times to reduce artificial gravitational shocking.
Total energy is conserved to better than $\sim$1\% of the satellite
internal potential energy during all simulations.

Our figures illustrate the results of simulations for satellites with masses $m = 6.5 \times 10^6 M_\odot$ to $m = 6.5 \times 10^8 M_\odot$  (where $m$
is the mass enclosed within 35 NFW\index{Navarro, Frenk, \& White (NFW) profile} scale radii, the radius out to
which particles were realized in the NFW distribution).
The scale radius r$_0$ was adjusted for each mass so that their density was the same with a base value of 0.86 kpc for the $6.5\times 10^6 M_\odot$ satellite. 
These satellites were allowed to  evolve for 8 Gyrs along orbits
with the same energy as a circular orbit at $R=25$~kpc in the host potential, but with angular momenta between 10\% and 90\% of that of a circular orbit (i.e. $L/L_{\rm circ}$=0.1 - 0.9) to contrast evolution on a near-circular to a highly-eccentric orbit.
%??\%  and??\% of mass was lost on the near-circular and highly-eccentric orbits during the course of each simulation.

\section{Basic principles of debris formation and evolution}
\label{sec:principles}

By definition, the tidal disruption of a satellite causes stars that used to be orbiting the satellite (which is in turn orbiting the parent galaxy) to become unbound and follow their own orbits around the parent galaxy.
%bound to and following the orbit of the satellite to 
%become unbound and follow new orbits.
Fig \ref{fig:evol} shows that, following disruption, the stars spread out through configuration space (i.e. they {\it phase-mix}\index{phase mixing}, roughly along the satellite's orbit) as the distribution of orbital properties in debris\index{tidal debris!orbital properties} corresponds to a distribution in orbital time-periods (described in Sect. \ref{sec:mixing}).
The rate at which the debris disperses reflects the scales over which debris orbital properties are distributed\index{tidal debris!scales of orbital properties} --- which are set by the nature of tidal disruption (described in Section \ref{sec:initial}).
The combination of tidal disruption and phase-mixing leads to phase-space morphologies for the debris that are primarily influenced by the mass of the satellite, its orbital path (which is in turn influenced by the parent galaxy potential), and the time since the debris became unbound. 
The scales of these morphologies can be broadly described by analytic formulae (Sect. \ref{sec:scales}) and represented by simple generative models (Sect. \ref{sec:models}).

\subsection{Debris spreading: phase-mixing\index{phase mixing}}
\label{sec:mixing}

The term {\it phase-mixing}\index{phase mixing} beautifully encapsulates the physics of the evolution of debris structures.
The nature of tidal disruption ensures that debris initially starts at the same {\it phase} (or position) along the orbit as the satellite from which it came, but with small offsets in its orbital properties (see Sect. \ref{sec:initial}).
These offsets mean that the typical orbital frequency and time-periods in the debris differ systematically from those of the satellite, and hence that the debris will increasingly trail or lead the satellite in orbital phase as time goes by, {\it mixing} along the orbit to form streams and shells.
%After enough time, the debris can become ((({\it fully phase-mixed}))), entirely filling the volume of phase-space available to the original orbit. In spherical

\subsubsection{Intuition from spherical potentials}

\begin{figure}[!t]
%\sidecaption
%\includegraphics[scale=0.35]{johnston_figs/fig3.eps}
\includegraphics[width=1.0\textwidth]{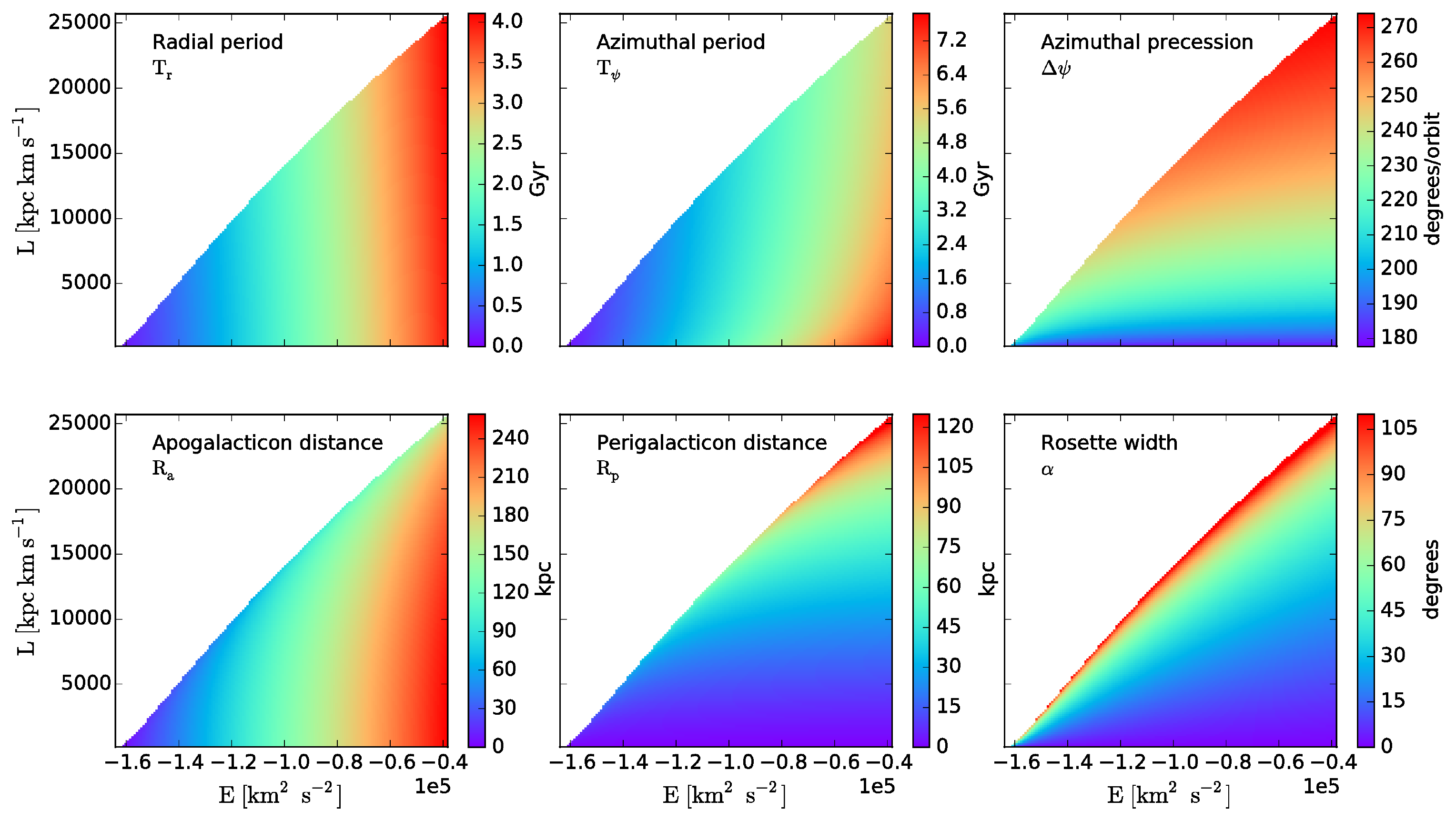}
\caption{Orbital properties in the spherical potential used in the N-body simulations as a function of energy (represented by the radius of a circular orbit of that energy) and angular momentum. The maximum angular momentum at a given energy is that of a circular orbit. The rosette width, $\alpha$, is the angle the orbit travels through during half a radial period, centered around apocenter. }
\label{fig:orbits}   
\end{figure}

Figure \ref{fig:orbits} illustrates how and why phase mixing occurs for the idealized case of orbits in the spherical NFW\index{Navarro, Frenk, \& White (NFW) profile} potential used in our simulations.
The top-left and upper-right-hand panels contour the radial time periods (time between successive apocenters or pericenters) and precession rate (angle between successive apocenters or pericenters) for orbits of a given energy and angular momentum per unit mass.  The energy of a particle in a circular orbit, in a given (spherically symmetric) potential, is completely determined by the (constant) radius of that orbit.  The orbital energy in Figure~\ref{fig:orbits} is represented by the radius of the circular orbit with that energy.
While the radial time periods depend much more strongly on energy than angular momentum, the precession angle depends more strongly on angular momentum, though to some extent on energy as well.
%on both.

\begin{figure}[!t]
%\sidecaption
%\includegraphics[scale=0.3]{johnston_figs/fig4.eps}
\includegraphics[width=0.9\textwidth]{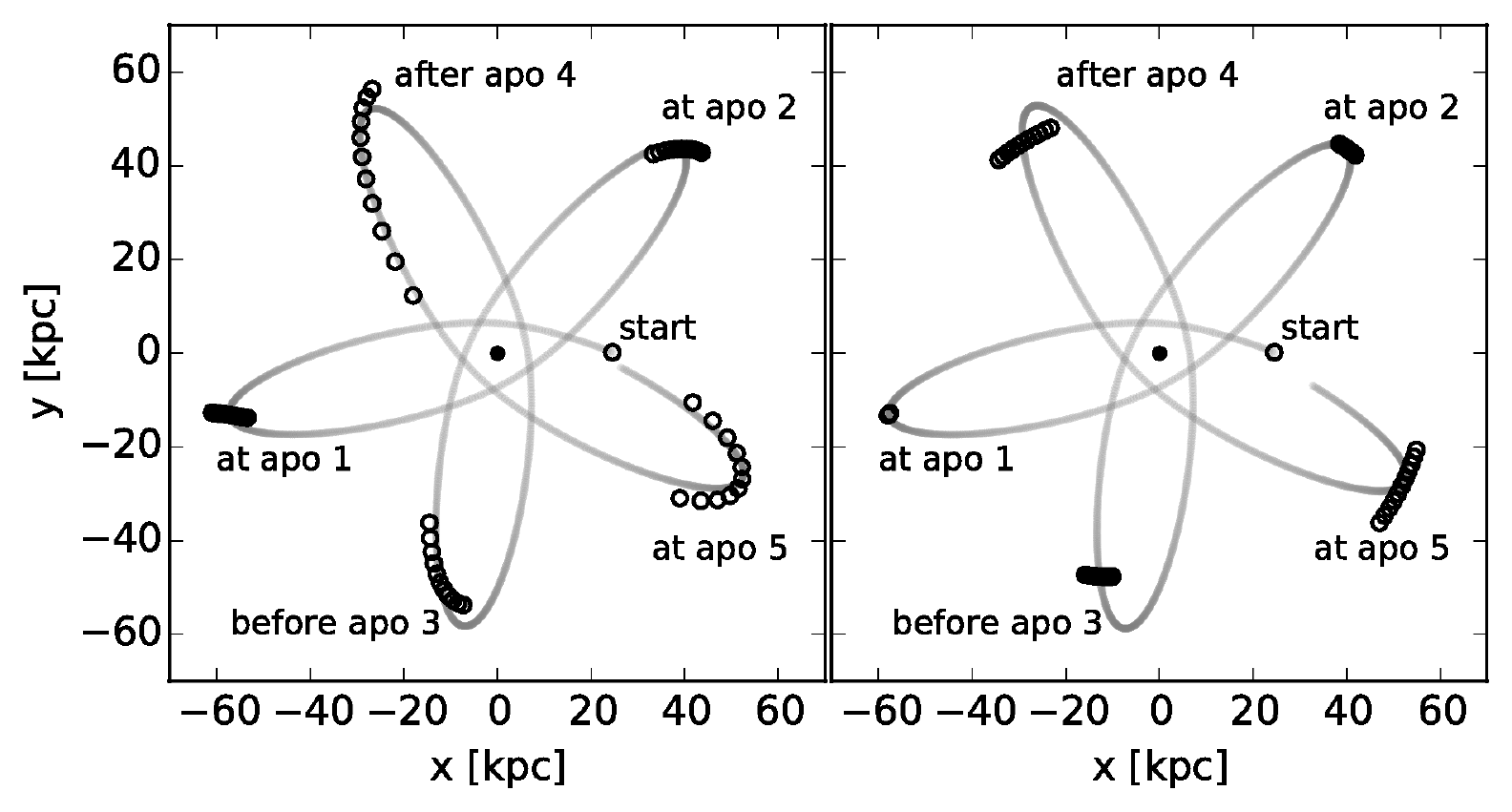}
\caption{Evolution of 11 particles that started at the same place along an orbit, but with a small spread in energy at fixed angular momentum (left hand panel) or angular momentum at fixed energy (right-hand panel).  The orbit of a galactic satellite is shown by the solid line, and the galaxy's center is shown by the solid dot.  The particles are shown at five subsequent times, one time for each petal of the orbit.  Note that the positions of the particles diverge in phase in just a few orbits.}
\label{fig:orbit_spread}   
\end{figure}

Figure \ref{fig:orbit_spread} shows how a set of moderately eccentric orbits that have a small range of orbital energies (right panel) or angular momenta (left panel) diverge in phase over just a few orbits. In particular note that: orbits with the same angular momenta but different energies spread in radial phase (i.e. positions along their oscillation in radius between apocenter and pericenter), while those with the same energies but different angular momenta spread in precession angle; the physical spread is much more striking for energy differences than for the equivalent angular momentum differences because the precession rate is a much weaker function of orbital properties than the orbital frequencies; and there is little dependence on orbital phase of these observations.

%The right-hand panels repeat the middle panels for a set of highly-eccentric orbits, revealing the same general trends.  However, in this case the physical extent due to differences in angular momenta can be comparable or even larger than the spread due to differences in energies, especially when the debris is at orbital apocenter.

\begin{figure}[!t]
%\sidecaption
%\includegraphics[scale=0.25]{johnston_figs/fig5.eps}
\includegraphics[width=1.0\textwidth]{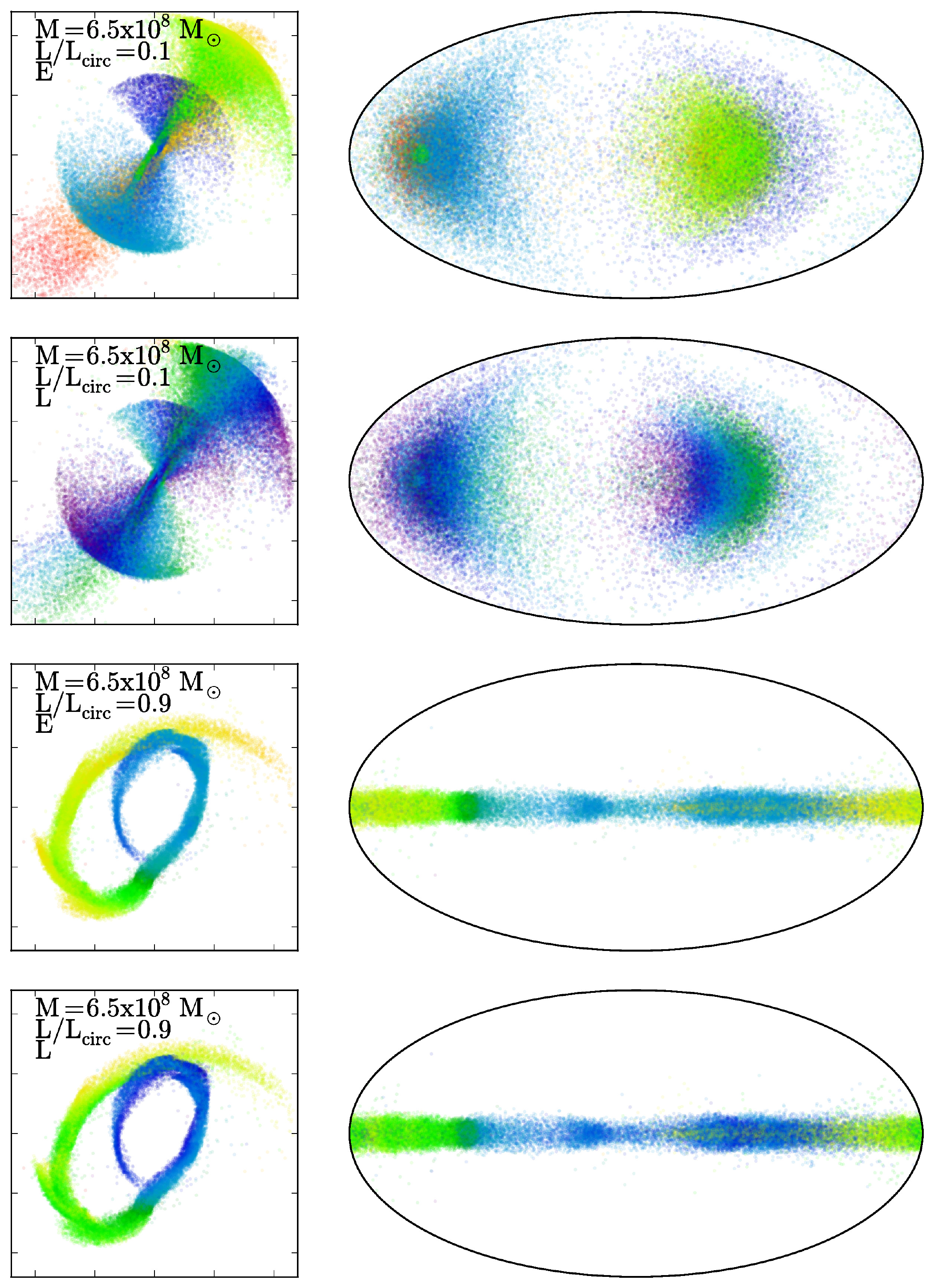}
\caption{Same as Fig. \ref{fig:structures} but color-coded by differences between the satellite and debris in energy (first and third rows) and angular momenta (second and fourth rows).}
\label{fig:ejsims}   
\end{figure}

Figure \ref{fig:ejsims} applies this intuition to one of our simulations, by color-coding particles with their differences in energy (first and third rows) and angular momentum (second and fourth rows) relative to the satellite from which they came. 
The sorting in energy along the orbital path and angular momentum around the orbital path is striking for both highly (top two panels) and mildly (lower two panels) eccentric orbits. 
It is also clear that the angular extent of the shells that form for the highly eccentric orbit is driven by differences in angular momenta, while the angular extent of streams is dominated by differences in energy.

Overall, these simple experiments give us
%allow us to gain 
an intuitive understanding of not only why phase-mixing occurs, but also 
%how 
the role that orbital properties play in determining whether debris is likely to form stream-like or shell-like morphologies.

\subsubsection{Action-angle formalism\index{action-angle formalism}}

The description in the previous section is useful for developing an intuitive understanding of debris evolution since it employs familiar orbital descriptors (i.e., energy and angular momentum) that can be written down analytically for any potential and any point in phase-space.
However, this description does not work for non-spherical potentials where angular momentum is not conserved and cannot be used to label orbits.
Debris evolution can be described elegantly for more general (but integrable --- see below) potentials using the action-angle variables of Hamiltonian dynamics.
A formal development of this description is given in \citet{helmi99a} \citep[and summarized in][]{bt08}.  We restrict ourselves here to a summary of the key ideas and equations.

Any point in phase-space can be located using the traditional spatial and velocity co-ordinates $({\bf x},{\bf v})$.
For regular (i.e. non-chaotic) orbits in {\it integrable} potentials, any phase space point can be equivalently labelled by conjugate variables $(\vec{\theta},\vec{J})$ where: (i) the actions, $\vec{J}$, are conserved orbital properties  which fully specify the path of the orbit through phase-space; and (ii) the angles, $\vec{\theta}$, represent the position along the orbit at which the point lies.
The actions and angles can be related through {\it Hamilton's equations}:
\begin{equation}
\label{eqn:hamilton}
\dot{\vec{J}}=-\frac{\partial{H}}{\partial{\vec{\theta}}}, \;\;\;  \dot{\vec{\theta}} = \frac{\partial{H}}{\partial{\vec{J}}},
\end{equation}
where the function $H$ (the {\it Hamiltonian}) is the energy of the orbit.
Since the actions are conserved along an orbit ($\dot{\vec{J}}=0$), $\partial{H}/\partial{\vec{\theta}} = 0$ and hence the Hamiltonian $H$ can only be a function of the actions, $H=H(\vec{J})$.
Then the Hamiltonian defines three unique orbital frequencies associated with each angle,
\begin{equation}
\label{eqn:freq}
\boldsymbol{\Omega} \equiv \dot{\vec{\theta}} = \frac{\partial H}{\partial \vec{J}},
\end{equation}
which are also conserved along the orbit since they are only functions of $\vec{J}$.
Equation (\ref{eqn:freq}) can be trivially integrated to give the position along the orbit at any time:
\begin{equation}
\label{eqn:angles}
{\vec{\theta}}={\vec{\theta_0}}+\boldsymbol{\Omega} t.
\end{equation}
Note that these Hamiltonian angles are not trivially related to the angular position along an orbit derived from position $(x,y,z)$:
equation (\ref{eqn:angles}) shows that they increase linearly with time at a rate that is not orbital-phase dependent (i.e. with constant frequency), unlike, e.g., the angular coordinate along an  eccentric orbit in a spherical potential, where the angular velocity is greater at pericenter than apocenter.

In this framework, the small range of orbital properties in tidal debris (illustrated with energy and angular momentum differences in Figure \ref{fig:orbits}) can be represented by small changes to the actions, $\Delta \vec{J}$, which in turn lead to a range in orbital frequencies,
\begin{equation}
\label{eqn:dfreq}
\Delta \Omega_i =\Delta \vec{J}  \frac{\partial{\Omega_i}}{\partial{\vec{J}}} = \Delta J_j \frac{\partial^2{H}}{\partial{J_j}\partial{J_i}}  .
\end{equation}
This equation can also be trivially integrated to describe debris spreading (i.e. phase-mixing)
\begin{equation}
\label{eqn:dangles}
\Delta \vec{\theta} = \Delta \vec{\theta_0} + \Delta \boldsymbol{\Omega} t.
\end{equation}

This action-angle description allows a re-statement of our understanding of the origins of debris morphology in terms of the properties of the {\it Hessian} matrix\index{Hessian matrix} --- ${\partial^2{H}}/{\partial{J_j}\partial{J_i}}$ in equation (\ref{eqn:dfreq}) --- which governs how orbital frequencies change in response to perturbations in the actions. This is a real symmetric matrix, which can be diagonalized and from which eigenvalues and eigenvectors can be derived. 
Assuming an isotropic distribution in actions (i.e. $\Delta J$ the same in every dimension --- not strictly true, see Sect. \ref{sec:initial} and Bovy, 2014) the eigenvectors of the Hessian define the principal directions (in angle) into which debris spreading occurs and the eigenvalues define the rates at which this occurs.
For realistic galactic potentials for which the Hessian has been derived, one eigenvalue has typically been found to dominate, indicating preferential spreading in just one dimension to form a tidal {\it stream}.

While action-angle variables offer an elegant description of debris dispersal, their versatility and applicability are not unlimited. 
Action-angles can be found for any spherical potential, but are only known for one family of non-spherical potentials: the triaxial St{\"a}ckel potentials\index{St{\"a}ckel potentials} \citep[see][]{bt08}.
Otherwise, they cannot generally be written down nor exactly derived numerically.
This means that the transformation from $(\vec{\theta},\vec{J})$ to the more familiar co-ordinates $(\vec{x},\vec{v})$ is non-trivial and hampers the development of a simple intuitive understanding of how they correspond to each other.
Moreover, this description breaks down entirely for chaotic regions of non-integrable potentials.
Approaches to tackle these limitations typically involve approximating general potentials in which the actions are unknown with ones in which they {\it are} known. Examples include representing a general potential using an expansion in St{\"a}ckel potentials \citep{sanders12} or calculating actions for an orbit in a general potential from the actions derived for a set of orbits in a slightly perturbed potential \citep{bovy14,sanders14}.

\subsection{Orbital properties of tidal debris\index{tidal debris!orbital properties}}
\label{sec:initial}

The previous section attributed the dispersal of stars stripped from satellite disruption to the range in orbital time-periods within the debris. This section examines what sets that range.

Suppose a satellite of mass $m_{\rm sat}$ is following a circular orbit of radius $R$ around a galaxy of mass $M_R$ enclosed within $R$, with speed $V_{\rm circ}=\sqrt{GM_R/R}$. 
Moving to a frame co-rotating with the orbit allows the definition of a time-independent {\it effective potential}\index{effective potential}, $\Phi_{\rm eff}$,
and the conserved {\it Jacobi integral}\index{Jacobi integral}, $E_J=v^2/2+\Phi_{\rm eff}$.
The {\it tidal radius}\index{tidal radius}, $r_{tide}$, at which the size of the satellite is limited by the gravitational influence of the parent can be estimated from the saddle points (the inner and outer {\it Lagrange points}) in $\Phi_{\rm eff}$  as:
\begin{equation}
\label{eqn:rtide}
r_{\rm tide}=\left(\frac{m_{\rm sat}}{3 M_R}\right)^{1/3} R
\end{equation}
\citep[see][section 8.3 for derivation]{bt08}, which also defines a limiting $E_{\rm J}$ for escape from the satellite.
However, since the escape of a star from the satellite depends on both its position and velocity, the tidal radius should not be considered as a solid boundary between bound and unbound stars. 
Moreover, it is not possible to define a strict tidal radius or $E_{\rm J}$ for satellites on non-circular orbits as there is no steadily-rotating frame in which the joint potential appears static. 
Nevertheless, numerical experiments have repeatedly demonstrated that the {\it tidal scale}\index{tidal scale},
\begin{equation}
s=\left(\frac{m_{\rm sat}}{M_R}\right)^{1/3} 
\end{equation}
suggested by equation (\ref{eqn:rtide}) captures much of the physics that creates debris distributions from satellites of a variety of masses and on a variety of orbits \citep{johnston98,helmi99a,eyre11,kuepper12,bovy14}.

\begin{figure}[!t]
%\sidecaption
%\includegraphics[scale=0.4]{johnston_figs/fig6.eps}
\includegraphics[width=0.95\textwidth]{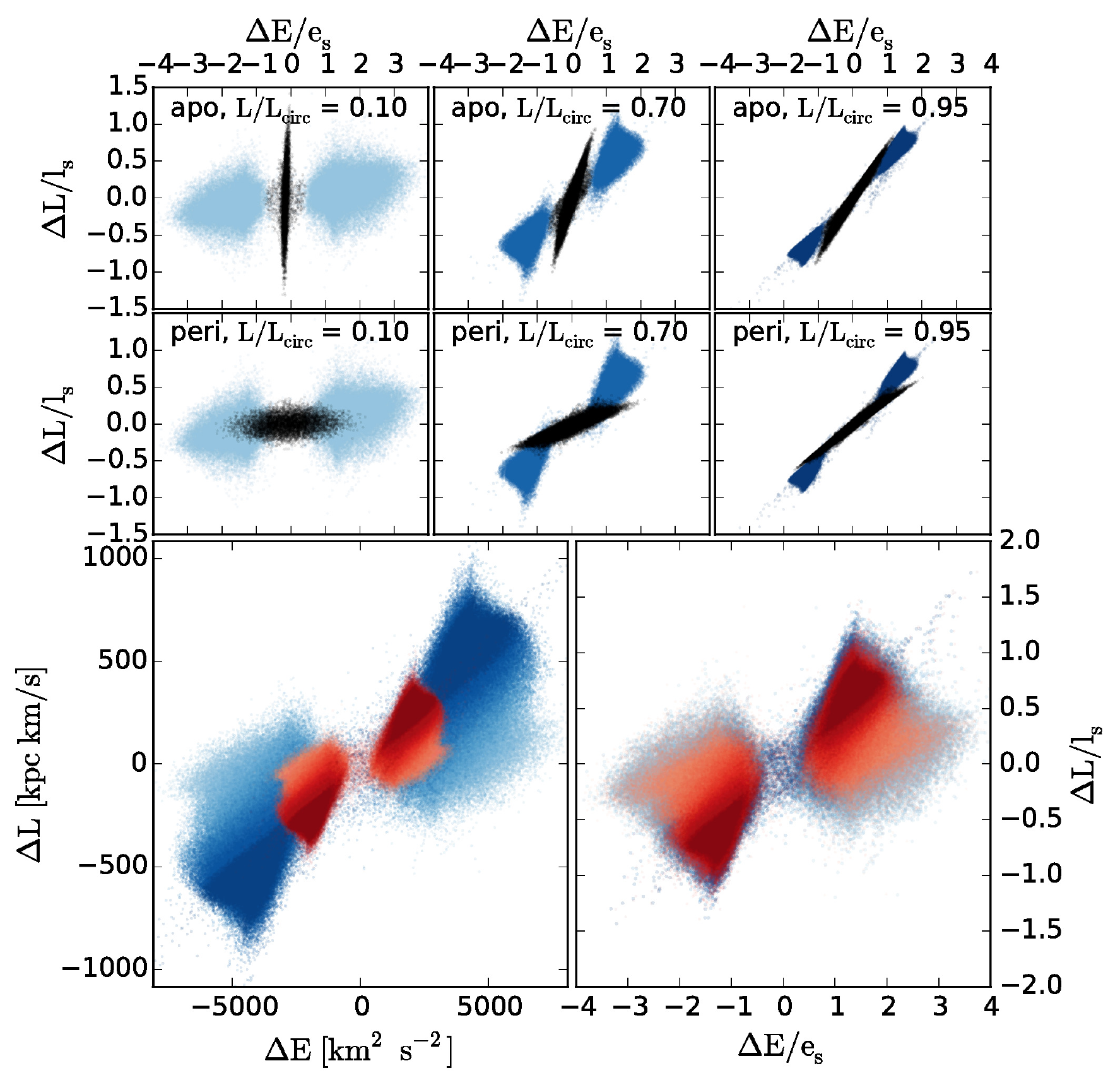}
\caption{Distribution in orbital properties for debris (blue) and satellite (black), 
%in orbital properties 
relative to the satellite's orbit at apocenter (top row) and pericenter (second row).
% for the 
The illustrated orbits are for a $6.5\times 10^7 M_\odot$ satellite on orbits with $L/L_{\rm circ}=0.1, 0.7, 0.95$.  Left panels show the most eccentric orbits, and the most circular are on the right, with each case color-coded with a particular shade of blue. 
%Note that the orientation of the orbit doesn't matter because we have used a spherical galactic potential.
The axes in the top 2 rows are scaled by the expected energy and angular momentum scales given in equations (\ref{eqn:scales}).  The blue points representing the debris are distributed similarly at apocenter and pericenter, for a given eccentricity.
The bottom panels combine the energy and angular momentum distributions for all eccentricities and for two different masses ($6.5\times 10^7 M_\odot$  in blue and $6.5\times 10^6 M_\odot$ in red) and plot them in physical units (bottom left) and scaled units (bottom right).  Material that is still bound to the satellite is not shown in the lower two panels.  Note that in scaled units (bottom right panel), the width of the leading/trailing distributions as well as the gap between them are similar across a wide range of eccentricities.}
\label{fig:orbprops}   
\end{figure}

Figure \ref{fig:orbprops} illustrates this understanding with plots of the orbital distributions\index{orbital distribution} produced in our numerical simulations (see section \ref{sec:nbody}), for satellites on orbits from highly eccentric to nearly circular (top panels, left to right). 
Each panel plots the energies ($\Delta E$) and  angular momenta ($\Delta L_z$)
%, radial action ($\Delta J_r$) and radial and azimuthal frequencies ($\Delta \Omega_r$ and $\Delta \Omega_\phi$) 
of escaped particles (in blue) relative to the satellite's own (i.e. at origin in each plot --- bound particles are shown in black).
The axes have been scaled by simple estimates for the energy and angular momenta
%, action and frequency 
ranges over which the particles are expected to be distributed,
\begin{equation}
\label{eqn:scales}
e_s=r_{\rm tide} \frac{d \Phi}{d R} \;\;\; {\rm and} \;\;\; l_s=s L_z,
\end{equation}
where $\Phi$ is the potential of the parent galaxy.

Each panel in Figure \ref{fig:orbprops} shows paired distributions of unbound particles corresponding to debris leading/trailing the satellite along its orbit at negative/positive values of $\Delta E$ corresponding to systematically larger/smaller frequencies and shorter/longer orbital time periods.
There is a distinct gap in orbital properties between the debris distributions for tidally stripped stars which is occupied by particles still bound to the satellite.
% occupy.
\citep[On the last pericentric passage, when all the particles become unbound the debris distribution will fill in this gap, see][] {johnston98}.
While the boundary between bound and unbound is not exact in configuration space, the separation between black and blue points in orbital properties is clear. 
The bottom panels contrast the debris distributions for all orbits (overplotted on each other) and two different satellite masses in physical (left-hand panel) and scaled (right-hand panel) units.
All together, the figures show that both the width of the  leading/trailing distributions and the gap between them are  similar in these scaled units across a wide variety of eccentricities.
Similar scaled distributions can be plotted for orbital actions and frequencies \citep[see, e.g.,][for a discussion]{bovy14}.

This uniformity in orbital properties for orbits with a range of eccentricities, when normalized with a single physical scaling, is one of the factors that enables the very simple descriptions of subsequent evolution outlined below in Sections \ref{sec:scales} and \ref{sec:models}.
Since the satellite is typically small, the global potential is dominated by the (nearly static) parent galaxy  and the debris orbital properties can be assumed not to evolve with time.
Of course this assumption is a simplification --- the gravitational influence of the satellite does not just turn off when a particle crosses the tidal radius but actually shapes the final distribution of debris orbital properties \citep[see][for some discussions of this]{choi09,gibbons14}.
%In particular, \citet{choi09} has shown that for larger satellite masses (above about 0.1\% of the parent galaxy, when $s\sim 0.1$) the s atellite can significantly alter the debris distribution, which in turn can drive orbital evolution of the satellite itself. Indeed, this interaction is apparent in the structure of the orbital distributions for our largest satellites --- the ridges along which particles align correspond to resonances between the particles' and the satellite's orbital frequencies.  For lower mass satellites, these interactions are not apparent both because the debris distribution is not broad enough to overlap these resonant regions and because the satellite itself is not massive enough to introduce a strong time-dependence to the global potential.

\subsection{Application: models of streams}
\subsubsection{Estimating physical scales in debris\index{tidal debris!physical scales}}
\label{sec:scales}

The understanding of distributions of orbital properties in tidal debris (Sect. \ref{sec:initial}) can be combined with the behavior of orbits (i.e. the azimuthal time periods, $T_\Psi$, for increasing the azimuthal angle by $2\pi$ and precession angles between turning points, $\Psi$, see Fig. \ref{fig:orbits} and Sect. \ref{sec:mixing}) to make simple predictions for the physical scales and time-evolution of the debris.
%For example, \citet{johnston01} give analytic formulae for the tidal and energy scales for the case of disruption around a galaxy with a perfectly flat rotation curve with circular velocity $v_{\rm circ}$, where the potential can be represented by $\Phi= v_{\rm circ}^2 \log(R)$:
%\begin{eqnarray}
%s &=&\left(\frac{m_{\rm sat}}{M_R}\right)^{1/3} = \left(\frac{G m_{\rm sat}}{v_{\rm circ}^2 R}\right)^{1/3} \\
%\epsilon &=&r_{\rm tide} \frac{d \Phi}{d R}= s v_{\rm circ}^2. 
%\end{eqnarray}
Exploiting the fact that orbital time periods in spherical potentials are largely independent of angular momenta (as illustrated in Figure \ref{fig:orbits}) the azimuthal time period for any orbit can be approximated by that of a circular orbit of the same energy $T_\Psi^{\rm circ}$. Then, the angular extent of the debris at time $t$ after disruption due to the characteristic energy scale over which it spreads ($\sim 4 e_s$ --- see equation [\ref{eqn:scales}] and Fig. \ref{fig:orbprops})  is of order:
\begin{equation}
\Psi_e = 4 e_s \left(\frac{1}{T_\Psi^{\rm circ}} \frac{\partial{T_\Psi^{\rm circ}}}{\partial{E}} \right) \frac{2 \pi t}{T_\Psi^{\rm circ}} 
\label{eqn:psie}
\end{equation}
Similarly, given the angle $\Psi$ between subsequent apocenters along the orbit of the parent satellite, the angular extent due to the spread in apocentric precession rates over the characteristic angular momentum scale ($\sim 4 l_s$) can be estimated as:
\begin{equation}
\Psi_l= 4 l_s  \frac{\partial{\Psi}}{\partial{L}}  \frac{t}{T_\Psi^{\rm circ}}.
\label{eqn:psil}
\end{equation}
Lastly, for a purely spherical potential, where the orbits are planar, the height $h$ of debris perpendicular to the orbital plane is set by the range of orbital inclinations available to debris escaping at the Lagrange (i.e. saddle) points in the effective potential with the characteristic ranges in energies and angular momenta. It has been shown to be of the same order as the tidal radius,
\begin{equation}
  h \sim s R \sim r_{\rm tide}.
 \end{equation}
 These simple formulae can be used to characterize the length $\Psi_e$ and width $\Psi_l$ in and height $h$ above the orbital plane for debris from a satellite of any mass disrupting along any orbit in any spherical potential \citep[e.g. as confirmed with N-body simulations in][]{johnston01}. 
Note that, since both $\Psi_e$ and $\Psi_l$ are proportional to time, the ratio $\Psi_e/\Psi_l$ is constant in time, and typically much greater than 1.  This explains the tendency for debris to form streams (as also suggested by the unequal eigenvalues of the Hessian in the action-angle description).
This ratio decreases for more eccentric orbits, which contributes towards the more shell-like appearance of debris in these cases (see section \ref{sec:young} for a more complete description).

%As discussed in Sect. \ref{sec:mixing}, $w$ will gradually increase with time due to the different turning point precession rates present in the debris, but for mildly eccentric orbits this simple expression is of the right order of magnitude  for dozens of orbits.  For more eccentric orbits,  the apparent angular extent of streams along the orbit $\Psi$ is actually limited by the preference for debris to spend most time at orbital apocenter and the increase in $w$ becomes dominant and clearly apparent as debris becomes inreasingly shell-like in morphology \citep[see][for a more complete discussion]{hendel15}.

\subsubsection{Generating predictions for density distributions along streams\index{density distribution!predictions}}
\label{sec:models}

The understanding of orbits and orbital distributions in debris can also be combined to build more detailed predictions for the full phase-space distribution along  tidal streams.
One approach is to first integrate only the orbit of the satellite and subsequently calculate at each phase along the orbit the centroid, width and density of debris that material tidally stripped over time must have under some assumptions for the mass-loss history of the disrupting object.
This was first done using approximate, analytic expressions for debris morphology, derived from energy and angular momenta considerations alone \citep[as outlined above and in more detail in][]{johnston98,johnston01}. 
\citet{helmi99a} instead followed how the density of a single packet of debris (i.e. unbound at the same point in time and at the same orbital phase) evolved using an action-angle formalism.
More recently, \citet{bovy14} and \citet{sanders14} have built models in action-angle space, where the full phase-space distribution of a stream can be predicted by calculating the expected offset in angles at a given orbital phase from a single orbit integration.
These latter models have the advantage over models built around energy and angular momenta in that they are also applicable to non spherical potentials and rely on a precise mapping rather than approximate scalings to calculate widths and offsets.
However, they do rely on having methods to calculate actions and angles in arbitrary potentials, which introduce an additional layer of both complications and approximations \citep[see][]{bovy14,sanders14}.

A second approach, slightly more computationally expensive but applicable to arbitrary potentials,
% is to use the debris orbital distributions derived from N-body simulations (e.g. as illustrated in Fig. \ref{fig:orbprops}) to provide the {\it initial conditions} (in phase-space) at the time when stars become unbound.
is to integrate the satellite orbit and, at each time-step, release a set of particles to represent the stars lost at that time.
At the point of release, the positions and velocites of the debris particles relative to the satellite are chosen to collectively reproduce the orbital distributions seen in full N-body simulations (e.g. as illustrated in Fig. \ref{fig:orbprops}).
The particles' subsequent orbital paths in the combined field of the host and satellite can then be simply calculated using test-particle integration.
At any point in time the integration can be stopped and the phase-space distribution of the particles be used to trace the resultant shells and streams.
This simple approach has been used both for debris modeling \citep{yoon11,kuepper12} and potential recovery \citep{varghese11,gibbons14}.

\section{Morphologies of individual debris structures\index{morphologies of debris structures} in observable co-ordinates}
\label{sec:observables}
%Louiville's theorem.
%?caustics? \cite{tremaine99}, 
A broad summary of the dependencies of debris structures can be drawn from  the physical models outlined in the previous section (and by inspection of Figure \ref{fig:evol}):

%\begin{description}
\begin{itemize}
\item{The {\it orbit}} of the progenitor satellite sets the large-scale morphology of the debris structure. 
\item{The {\it mass}} of the satellite sets the scales over which debris is distributed. 
\item{The {\it time}} since the satellite starts losing stars determines the degree to which the debris is phase-mixed. 
%\item{The {\it potential}} of the parent determines both the exact properties of the satellite orbit and the nature of debris dispersal. Most obviously, in spherical potentials orbits are planar and debris spreads only in two spatial dimensions; in non-spherical potentials both the orbits an debris spreading explores all three dpatial dimensions (\cite[helmi99]).
\end{itemize}
%\end{description}
 Armed with these descriptions we can now go on to interpret observations of debris structures in terms of the history of the progenitor satellite. There is a rich literature with exact models (typically based on N-body simulations) of individual structures \citep{johnston95,velazquez95,helmi01,penarrubia05}. Here we instead discuss in principle how and why these models are uniquely sensitive to progenitor properties.

Note that the potential of the parent galaxy also affects debris properties, so debris can also be used to constrain the mass distribution in the Milky Way. This will be discussed in Chapter 7.

\subsection{Young debris\index{young debris}}
\label{sec:young}
As debris ages, it spreads further and further apart in orbital phase away 
from the progenitor along its orbit, overall  decreasing in density over many orbits. 
The debris is considered {\it fully phase-mixed}\index{phase mixing!full} once it fills the configuration-space volume defined by the progenitor's orbit.
For example, for an eccentric orbit in a spherical potential, fully phase-mixed debris would be a spread over a (near-planar) annulus with inner and outer radii corresponding to the pericenter and apocenter of the parent satellite's orbit.
For a loop orbit in a potential where the orbital plane precesses, the debris would fill a three-dimensional donut shape.
For a given orbit, debris becomes fully phase-mixed much more rapidly for more massive progenitor objects.

The term ``young debris'' is used to refer to structures that have {\it not} had time to fully phase-mix and are apparent as distinct spatial overdensities in star-count maps, such as the Sagittarius and Orphan streams, and the GD-1 and Pal 5 globular cluster streams.
These have been found at distances of 10 kpc to 100 kpc from the Sun where the background stellar density is low enough for such low surface brightness features to be apparent
 and orbital timescales sufficiently long for mixing not to have proceeded far during the lifetime of the Galaxy.

As illustrated in Figure \ref{fig:structures}, streams result from satellite destruction along mildly eccentric orbits. 
In these cases, the spreading due to differences in turning point precession ($\Psi_l$, see equation [\ref{eqn:psil}]) is much smaller than the initial angular width of the stream (of order $s R \sim r_{\rm tide}$) for many orbits and always less than the spreading along the orbit due to differences in orbital time ($\Psi_e$, see equation [\ref{eqn:psie}]).
Hence the {\it width} of a stream is an indication of the progenitor mass. 
Once the mass is known, the age of the debris (or time since the satellite first started losing stars) can be estimated from the {\it length} of the streams.

Shells result from destruction along more eccentric orbits.
There have been extensive studies of the properties and interpretation of shell systems seen around external galaxies \citep{quinn84,sanderson13}, though these have largely concentrated on structures formed along almost radial orbits.
%In contrast, the devlopment of an analogous understanding of clouds (i.e. shells seen form an internal perspective) is in its infancy.
The transition between conditions that produce stream-like or shell-like morphologies\index{morphologies of debris structures!stream-like vs. shell-like} primarily depends on orbital eccentricity
\citep[e.g.][found this to occur on orbits where $L/L_{\rm circ} \sim 0.3-0.5$]{johnston08},
but also depends on satellite mass and time since disruption.
Both stream-like and shell-like morphologies can occur in the same structure in this transition region. 
For a given mass and orbit, streams are most  apparent in the early stages of evolution as the spreading along the orbit that produces them (i.e. as estimated by $\Psi_e$ in equation [\ref{eqn:psie}])  typically  occurs more rapidly than the differential precession of the apocenters that gives rise to shells (i.e. as estimated by $\Psi_l$ in equation [\ref{eqn:psil}]).
However, along the more eccentric orbits,  the rapid passage of debris through pericenter  significantly reduces the density at these  orbital phases and the apparent angular extent of contiguous streams can be effectively limited by the angular size subtended by the region around the apocenter of the orbit where debris spends most of its time \citep[estimated in][as the angle centered on apocenter where the satellite orbit spends half of its time, $\alpha$ in Figure \ref{fig:orbits}]{hendel15}.
Under these circumstances, a single disruption event can produce what appear to be distinct structures, centered at two or more orbital apocenters with orbital precession becoming the dominant effect that dictates the apparent angular spread around each one.
The time at which these structures start taking on shell-like characteristics can be estimated by finding when $\Psi_l $ is greater than $\alpha$.
\citep[See also][for an independent discussion of these effects]{amorisco14}.

\subsection{Fully phase-mixed debris\index{phase mixing!full}}
\label{sec:old}
% ???TO DO ???? Caustics here? Helmi & White? Tremaine? Sanderson?
%As debris ages, it spreads further and further apart in orbital phase away 
%from the progenitor along its orbit, overall decreasing in density. 
%The debris is considered fully phase-mixed once it fills the configuation-space volume defined by the progenitor's orbit.
%For a given orbit, this occurs after a much shorter time for debris from more massive progenitor objects.

Despite the low density and lack of spatial coherence of fully phase-mixed debris, its presence can often still be detected.
For example, Liouville's theorem\index{Liouville's theorem} states that the flow of points through phase-space is {\it incompressible}: that the phase-space density of debris remains constant in time even as it evolves to form streams and shells.
An inevitable consquence is that  debris must become locally more concentrated in velocity space even as it becomes more diffuse in configuration space \citep{helmi99a}.
Hence, signatures of the disruption of satellites can remain apparent in catalogues of stellar velocities even if no  spatial structures remain detectable. 
This idea was first conjectured by Olin Eggen \citep[for a summary see][]{eggen87}, who proposed that nearby moving groups of stars could be the remnant of long-dead star clusters.
In Chapter 5 the more recent work on velocity substructure in the stellar halo \citep[e.g.][]{xue11,schlaufman09} is discussed in more detail.

Surveys with full phase-space information can exploit the fact that the orbital properties of the debris (e.g. their actions or frequencies) remain constant in a static potential.
\citet{helmi99b} were the first to apply this idea to data using the Hipparcos\index{Hipparcos mission} catalogue of proper motions and parallaxes to derive angular momenta and estimate energies for giant stars within 1kpc of the Sun.
They found $\sim$10\% of the stars in their sample to be clumped in orbital-property space and concluded that 
10\% of the local halo must be formed from an object similar to the dwarf galaxy Sagittarius being disrupted in the distant past.

Once larger samples with more accurate orbital properties are found in near-future surveys, 
\citet{gomez10a} have shown how the ages of these structures might also be determined by looking for substructure in orbital properties within a group \citep[see also][]{mcmillan08}.
To develop some intuition for this idea, consider the (incorrect) model of debris spreading exactly along a single orbit.
In the early stages of mixing the local volume will contain just one wrap of the debris stream at one orbital phase.
As the debris spreads, the local volume will gradually fill with more and more wraps.
The spreading itself is caused by the debris having a range of orbital properties. 
Hence, each wrap of the debris within the local volume will have different orbital properties and the degree and spacing of substructure within the orbital properties of the group gives an estimate of the age.

The beauty of using orbital properties to identify debris members is that stars can be connected even if they have no clear association in velocity or configuration space alone. 
For example, looking to the future, combining a distance indicator with {\it Gaia}'s\index{Gaia} assessment of proper motions and radial velocities suggests that satellite remnants throughout the inner 100 kpc of the Milky Way might be identified using this method even if their members are spread on disparate planes and a wide range of radii.

\section{Debris in a cosmological context\index{tidal debris!in cosmological context}: modeling and interpreting properties of stellar halos}
\label{sec:cosmology}
Within the current cosmological paradigm for structure formation, galaxies are thought to form, at least in part, {\it hierarchically}\index{hierarchical galaxy formation}, with small galaxies forming first within small dark matter halos and gradually agglomerating to form larger galaxies within larger dark matter halos.
Gas from this agglomeration process can dissipate and fall towards the center of the main halo to form a new generation of stars in the combined object.
Unlike the gas, stellar orbits are dissipationless, so, once stripped, the stellar populations of infalling galaxies can be left behind orbiting in the halo of the galaxy.
Of course, the orbits of infalling galaxies are affected by dynamical friction\index{dynamical friction}.
For those that are more massive than a few percent of the parent this can lead to significant evolution within a few orbital periods so stars from these objects can potentially make a minor addition to the central stellar galactic components that are forming from the {\it in situ}\index{in situ formation} gas (i.e. spheroid or disk, as seen in the hydrodynamic simulations of  \citet{abadi06}).
In contrast to the spheroid and disk, {\it stellar halos} are a great place to look for stars that have been accreted from other objects.  Current models and data favor a picture where a significant fraction --- and possibly all --- of the stars in the halo originally formed in other objects.

The accreted nature of at least a significant fraction of stars in galactic halos suggests their observed properties can be used to address a number of questions.
Cosmological parameters dictate the nature of hierarchical clustering: the frequency and epoch of infall of dark matter halos of different mass-scales and their orbital properties.
In other words, the cosmology sets the parameters discussed in Sect \ref{sec:observables} that fully specify the type and number of  tidal disruption events that have occurred.
Hence, adopting a given cosmology (with the addition of some assumptions for how stars occupy different dark matter halos) leads to specific expectations for the level of substructure due to accretion within stellar halos (Sect. \ref{sec:halos}) as well as chemical trends between substructure and field stars (Sect. \ref{sec:chem}).
These understandings can be exploited now and in the future to understand to what extent the distribution on the Milky Way's stellar halo matches our expectations (Sect. \ref{sec:stats}), and what we can learn about our accretion history from the extent and properties of substructure (Sect. \ref{sec:history}).
In turn --- identifying stars that were originally formed in other objects at much higher redshift offers a unique perspective on the stellar populations in these galactic progenitors (Sect. \ref{sec:higherz}).

\subsection{Cosmological simulations of stellar halo formation\index{cosmological simulations}}
\label{sec:halos}

Stellar halos are hard to observe because they contain a tiny fraction of stars (and an even tinier fraction of the total mass) in a galaxy, spread out over a large volume.
Moreover, surveys need to be sensitive to even smaller fractions of stars to learn about substructure within these halos, and the presence of this substructure makes it challenging to characterize the global characteristics of the halo itself.
Early discussions were largely restricted to global properties and formations scenarios (e.g. see classic works by \citealt{els62} and \citealt{sz78}) extrapolated from either local studies of high-velocity stars, larger volume surveys with tracer populations where distances could be estimated, or pencil beam surveys.
The field has been revolutionized in the last two decades with the emergence of large-scale stellar surveys covering a significant fraction of the sky, such as the Sloan Digital Sky Survey\index{Sloan Digital Sky Survey (SDSS)} \citep{abazajian03} and the Two Micron All Sky Survey\index{Two Micron All Sky Survey (2MASS)} \citep[e.g.][]{majewski03}.

Models of stellar halos\index{stellar halos!models of} face the same challenges as the observations in resolving such a tiny component of the galaxy, as well as substructure within it.
One approach is to restrict attention to only stars that have been accreted from other systems and hence avoid the need to follow gas physics and ongoing star formation explicitly. 
\citet{bullock01} presented a first attempt by combining a semi-analytic, generative model of tidal disruption events \citep[developed in][and outlined in Section \ref{sec:models} above]{johnston98} with merger histories for Milky-Way like galaxies predicted in a cosmological context using the Press-Schechter formalism \citep{lacey93}. 
These models showed abundant substructure in the model halos, which was coincidently being mapped by SDSS in the real stellar halo.
\citet{bullock05} made the next step to a more sophisticated approach by replacing the simple generative model of each disruption event with an N-body simulation to represent the dark matter evolution of the infalling object. 
Stars were ``painted'' onto the purely dark matter satellites by assigning a variable weight or mass-to-light ratio to each N-body particle.
The weights were chosen  in such a way that the properties of the latest infalling objects matched the internal spatial and velocity distributions observed for stars in nearby dwarf galaxies.
The \citet{bullock05} models were limited in that the  parent galaxy was represented by (slowly evolving) analytic functions and the results of the separate simulations were superposed only at the present time to make a stellar halo model.
Hence they did not include either the influence of the accreting objects on the parent or on each other. 
Nor did they represent the cosmological context (such as preferential infall along filaments or global tidal influences due to nearby neighboring structures).
Computational power is now such that full self-consistent N-body simulations of the formation of a Milky Way sized dark-matter halo are sufficiently resolved to address this limitation and there are now several example of stellar halo models made by painting stars onto these more realistic backdrops \citep{delucia08,cooper10,lowing15}.

Of course, the ultimate goal is to simultaneously build a model of all components of a galaxy (dark matter along with stellar and gaseous components) using cosmological hydrodynamical simulations of structure formation capable of following the baryonic as well as dark matter physics.
There are several examples already in the literature where the stellar halo components have been  resolved in these models and their characteristics and history discussed \citep{abadi06,zolotov09,font11,tissera13}.
In particular, these models drop the assumption that all stars in the halo come from accretion events and allow an exploration of how much of the stellar halo might have instead formed {\it in situ}.
However, the properties of the stellar halos vary systematically between the simulations,  as prescriptions for star formation and feedback also vary, so the results at this point seem indicative rather than conclusive.

For the remainder of this section we concentrate solely on the accreted component of the stellar halo and illustrate some general results common to all the models using the \citet{bullock05} simulations.

\subsection{General results of cosmological accretion models}

\subsubsection{Accreted phase-space structure in halos}
\label{sec:xv}

\begin{figure}[!t]
%\sidecaption
%\includegraphics[scale=0.4]{johnston_figs/bj05_halos.eps}
\includegraphics[width=0.95\textwidth]{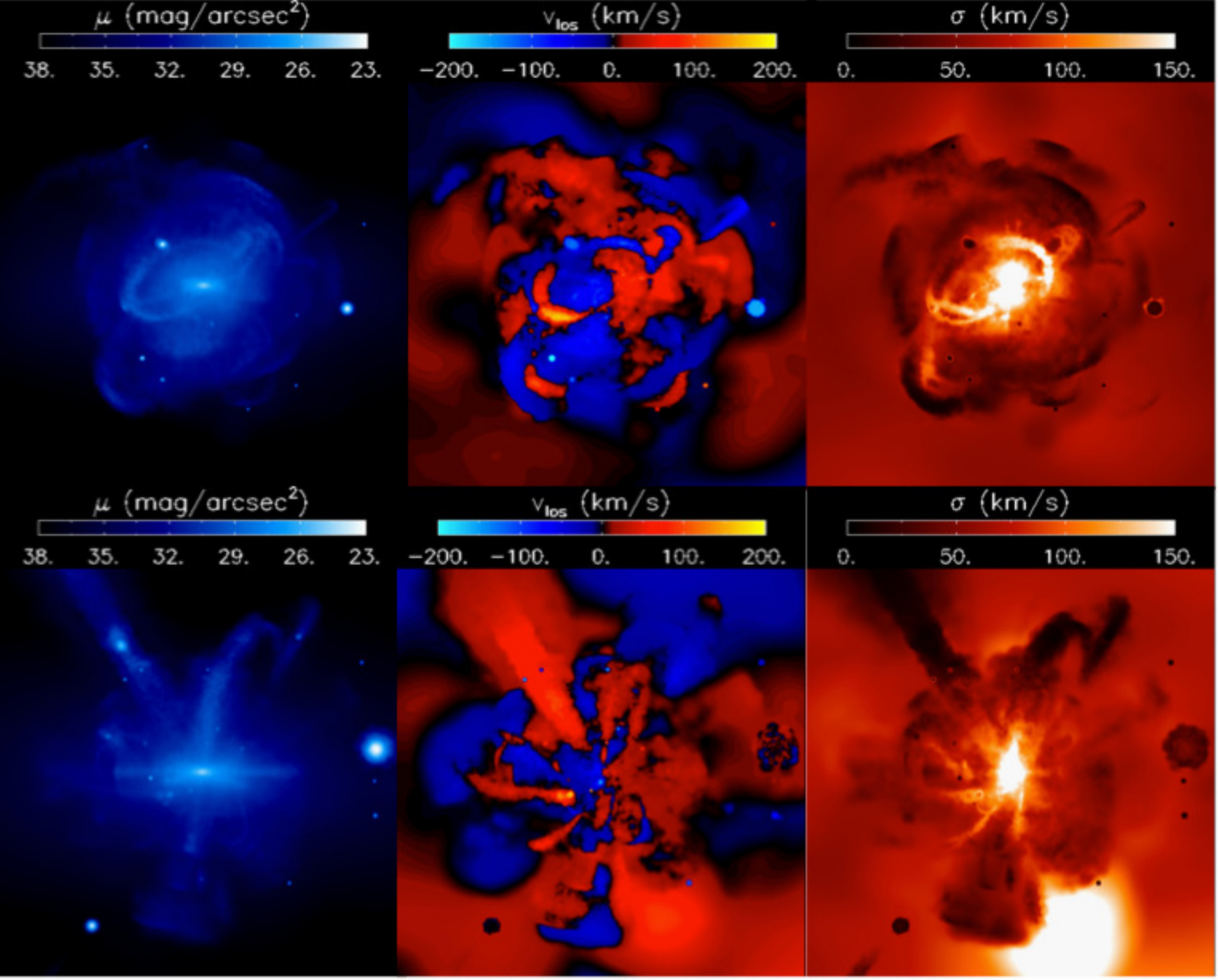}
\caption{Surface brightness (left), line-of-sight velocity (middle) and velocity dispersion (right) from external views of two stellar halo models built entirely from accretion events drawn from a merger history consistent with our current expectations \citep[from][]{bullock05,johnston08}. Each box is 300 kpc on a side. Only the stellar halo component is shown. Image credit: Sanjib Sharma.}
\label{fig:halo}   
\end{figure}

Figure \ref{fig:halo} shows one of the \citet{bullock05} purely-accreted stellar halo models from external viewpoints, in space and velocity.
The simplicity of the model permits sensitivity to both small as well as low surface brightness substructures within the halo.
Generic features of this (and other) models are: a smooth, fully phase-mixed inner region; abundant substructure in the outer parts, detectable both in space and velocity; and an increasing prevalence of the substructure with Galactocentric radius.
Beyond the phase-mixed inner regions, such model halos typically appear dominated by a handful of striking shells and streams, with shells tending to be more prevalent at the largest distances and streams in the intermediate parts \citep{johnston08}.

\begin{figure}[!t]
%\sidecaption
%\includegraphics[scale=0.4]{johnston_figs/bj05_special.eps}
\includegraphics[width=0.95\textwidth]{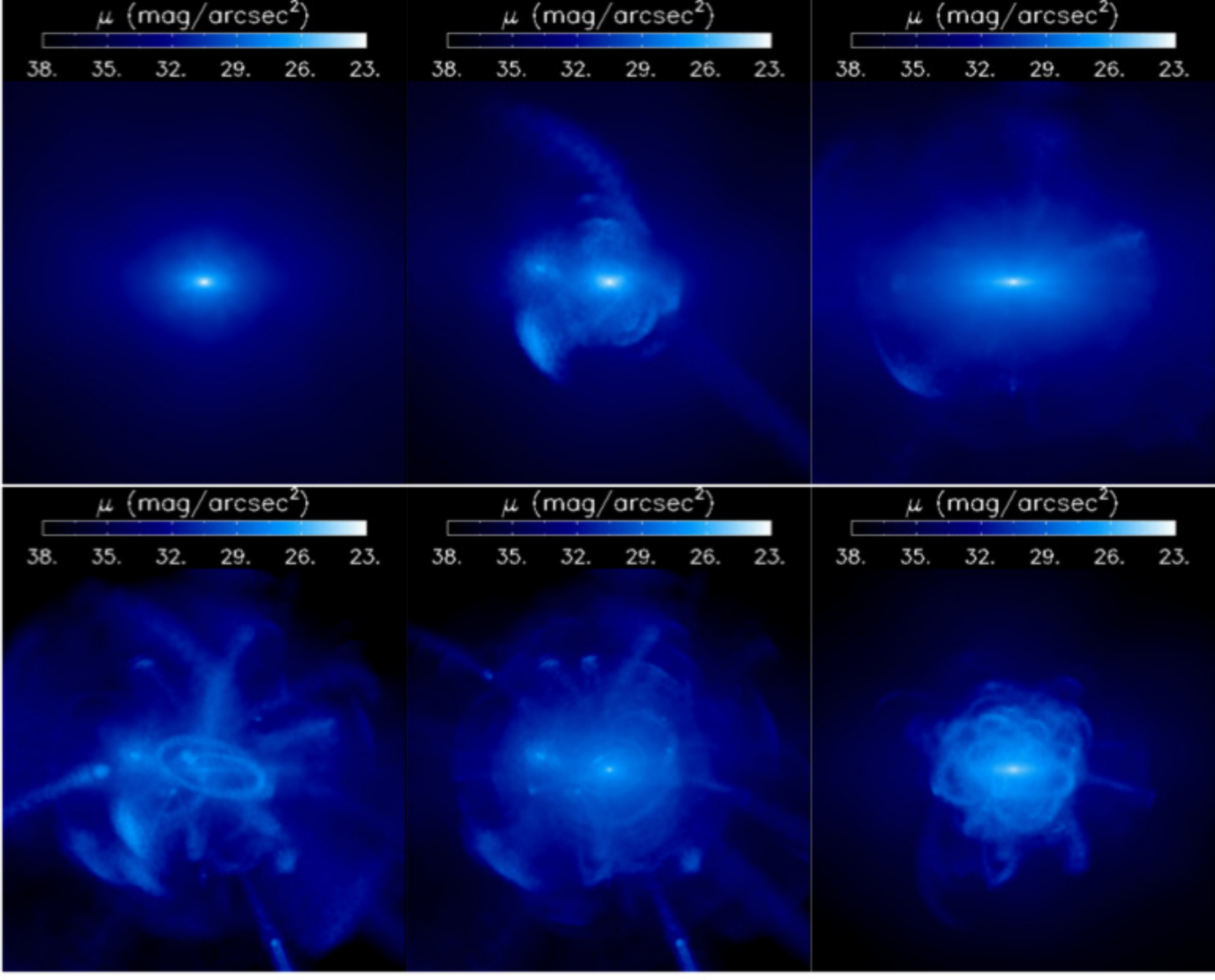}
\caption{Surface brightness projections for stellar halos with the same total luminosity, but merger histories that were artificially constrained to be dominated by different types of accretion events  \citep[following][]{johnston08}.
In the left panels, the events had the same luminosity and orbit distributions, but were either all accreted a long time ago (upper panel) or recently (lower panel).
In the middle panels, the events had the same accretion time and orbit distributions, but were either all of high (upper panel) or low (lower panel) luminosity.
In the right panels, the events had the same accretion time and luminosity distributions, but were either all on near-radial (upper panel) or near-circular (lower panel) orbits.
Image credit: Sanjib Sharma.}
\label{fig:althalos}   
\end{figure}

These generic features can broadly be explained in the context of the current cosmological expectations which suggest a history for the Milky Way where: (i) the majority of accretion events occurred more than 7-8 Gyrs ago; (ii) the events had a range of luminosities associated with them; and (iii) the accretions occurred on a mixture of orbits.
Figure \ref{fig:althalos} illustrate this with external views of model halos, constructed by \citet{johnston08}, that are instead built entirely from: (i) ancient or recent accretion events (left panels); (ii) high or low luminosity events (middle panels); and (iii) events evolving on high or low eccentricity orbits (right panels).
The consequences of these differences are obvious with a simple visual comparison of the panels 
and can be easily explained with the physical intuition developed in Sect. \ref{sec:principles} and Sect. \ref{sec:observables}: 
younger/older halos are more/less substructured because of the time available for phase-mixing; larger/smaller substructures correspond to higher/lower luminosity events because the total mass sets the tidal scales at distruption; and the orbit distribution dictates the debris morphology because more/less eccentric orbits tend to produce shells/streams.

\subsubsection{Accreted stellar populations in halos\index{stellar halos!stellar populations in}}
\label{sec:chem}

\begin{figure}[t]
%\sidecaption
%\includegraphics[scale=0.4]{johnston_figs/bj05_chem.eps}
\includegraphics[width=0.95\textwidth]{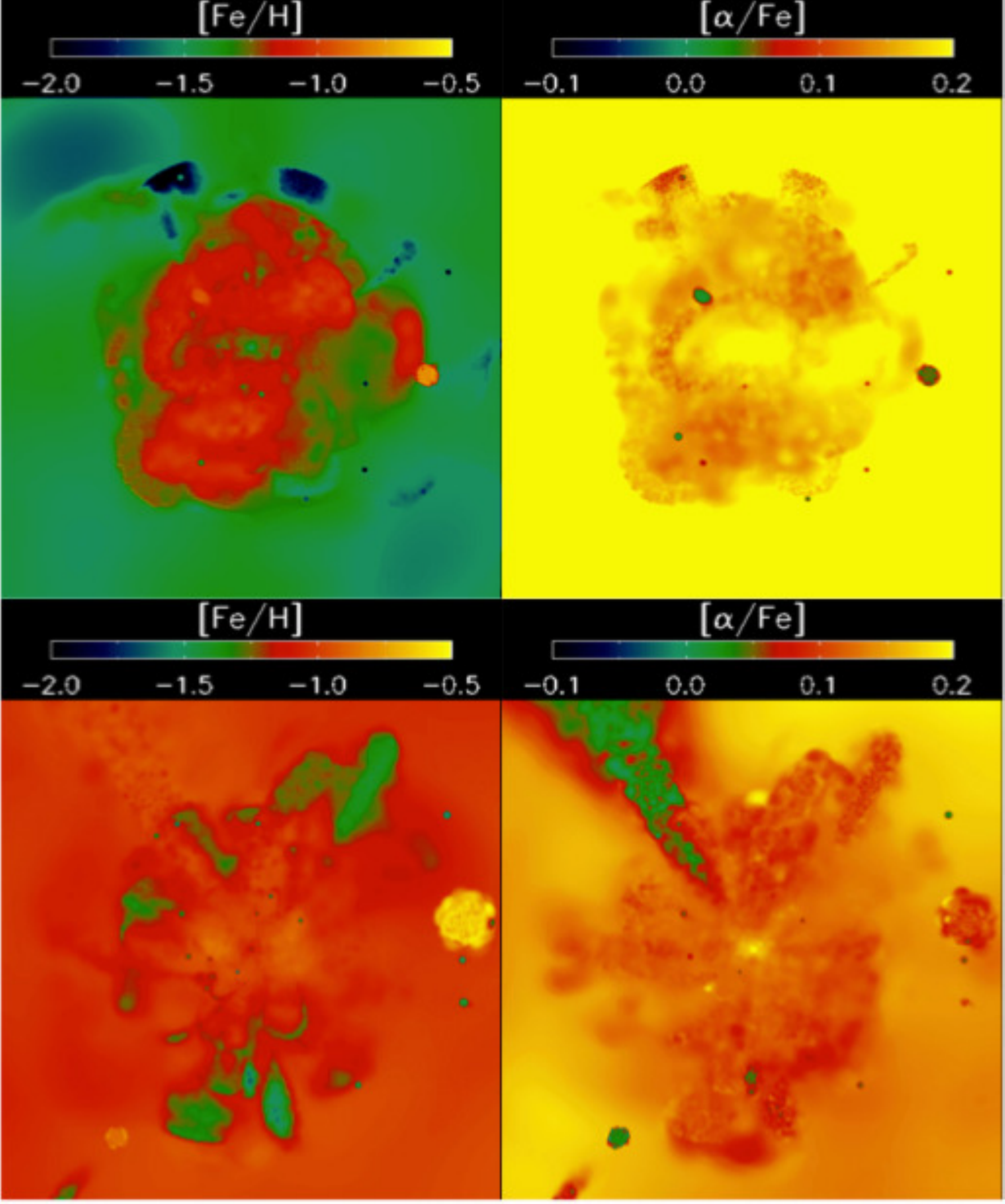}
\caption{Average [Fe/H] and [$\alpha$/Fe] projected along the line-of-sight for the two stellar halo models shown in Figure \ref{fig:halo}
\citep[using chemical model developed in][]{robertson05,font06}. Image credit: Sanjib Sharma.}
\label{fig:chem}   
\end{figure}

Figure \ref{fig:chem} shows an alternative visualization of the model stellar halos shown in Figure \ref{fig:halo}, but with the grid points color coded by the average metallicity and [$\alpha$/Fe] abundance ratio along the line-of-sight.
These were derived by assigning a simple star formation history to each infalling dwarf and running a leaky-accreting box model to estimate the associated chemical evolution \citep{robertson05,font06}.
The parameters of the chemical evolution models were tuned to reproduce known properties of dwarf galaxies in the Local Group today --- the observations that more luminous dwarfs tend to be more metal rich \citep[e.g.][]{grebel03}, and that all nearby objects contain $\alpha$-poor populations \citep[e.g.][]{venn04}.
Star formation was truncated in each dwarf at the time when it was accreted onto the Milky Way, as might be expected given that dwarfs near larger galaxies are observed to be quenched relative to their field counterparts \citep{grebel03,geha12}.
The last attribute of the model effectively means that the cosmological framework also influences the nature of the stellar populations in accreted components of galactic stellar halos, setting the characteristic time over which star formation can occur.

In the left-hand panels of Figure \ref{fig:chem} the mass-metallicity relation is apparent, as the largest and most dominant debris structures tend to be the most metal rich.
The influence of the cosmological background is apparent in the right-hand panels; the smooth stellar halo component is $\alpha$-enhanced relative to both surviving satellites and the brighter debris features.
Physically this trend can be attributed to the relative delay expected following star formation of Type Ia Supernovae (SNe) compared to Type II SNe.
The progenitors of Type II SNe are massive stars, whose deaths produce both iron and $\alpha$-elements, within a few million years of a star formation event.
In contrast, Type Ia SNe, producing mainly iron, arise from the explosion of an accreting white dwarf star  --- objects which will not form for hundreds of millions of years after a star formation event.
Hence, the oldest stellar populations in infalling dwarf galaxies are not expected to have been polluted by SNe Type Ia and should be rich in $\alpha$ elements, while younger populations will be relatively $\alpha$-poor.
In our accreted halo model, the smooth, fully phase-mixed portion of the halo comes from early infalling objects that do not have a chance to ever make the younger populations.
In contrast, these younger populations are apparent in more recently destroyed objects or surviving satellites that generally fell in even more recently.
\citet{zolotov10} points to analogous trends in $\alpha$ element patterns  when contrasting hydrodynamic simulations of the formation of galaxies with differing merger histories. 

Distinctions between the chemical properties of field stars in the halo and satellite galaxies that have been known about for some time \citep{unavane96,venn04} can naturally be explained within this cosmological context \citep{robertson05,font06}. 
Moreover, studies of stellar populations in the satellites, stellar halo and debris around M31 suggest this scenario can also be applied to understand variations there \citep[see, e.g.,][and Chapter 8]{font08,gilbert09}.

\subsection{Implications and applications}
\subsubsection{Statistical comparisons with observations}
\label{sec:stats}

As discussed in the previous section, combining our cosmological picture of how structures form in the Universe with tidal disruption and chemical evolution models leads to some specific expectations for phase-space and stellar population characteristics of debris structures as well as some  general trends. 
Both the characteristics and trends are broadly consistent with current observational surveys.
However, the stochastic nature of hierarchical structure formation means that there is large variation about the average properties among the stellar halo models produced and more quantitative comparisons employing a statistical approach are just starting.

The average spatial structure of the stellar halo, as well as the level of substructure within it, can be assessed using large scale photometric catalogues of stars.
For example: \citet{bell08} fitted triaxial, power-law models to star counts of main-sequence turnoff stars selected from SDSS and also quantified the level of deviations around these smooth models; and \citet{sharma10} exploited the distinct colors of metal-rich, evolved stars in the 2MASS filters \citep[following][]{majewski03} to select distant M-giant stars and ran a group-finding algorithm on the selection in the space defined by their angular position and apparent magnitude \citep[from][]{sharma09}.
In both studies, the analyses were repeated on equivalent synthetic stellar samples generated from the simulated stellar halos of \citet{bullock05}. 
The results (both numbers and scales of groups and level of deviations from a smooth model) varied significantly between the eleven different simulated stellar halos, with the results from the analysis of real data sitting within this spread.
While this agreement is encouraging, when \citet{helmi11} repeated the \citet{bell08}  analysis on the \citet{cooper10} model stellar halos (which were derived by ``painting'' stars in the ``Aquarius'' self-consistent dark matter simulations), they found systematically larger deviations from a smooth background than the prior work at a level that was inconsistent with the observations.
Recent work by \citet{bailin14} contrasting simulations with and without Galactic disk components suggest that this inconsistency with both observations and the \citet{bullock05} work might be attributed to the Aquarius simulations lacking the extra potential structure due to the disk.

Following photometric surveys with spectroscopic surveys allows assessments of the level of spatially correlated velocity substructure \citep[e.g., using K-giants in the {\it Spaghetti Survey}, metal-poor MSTO stars from SEGUE, or BHBs from SDSS, see][]{starkenburg09,schlaufman09,xue11}.
Currently, comparisons to models seem consistent, but again not conclusive \citep{xue11}.

Stellar populations in accreted halos are expected to exhibit spatial variation. These spatial variations
%Stellar population variation in space is also an expectation for accreted halos and these 
have been found photometrically  \citep[by looking at the ratio of MSTO to BHB stars across the sky in the SDSS catalogue, see][]{bell10} and spectroscopically \citep{schlaufman12}.
In particular, \citet{schlaufman11,schlaufman12} looked at the Fe- and $\alpha-$ element abundances of the velocity substructures they had found, and concluded that they tended to be chemically distinct from the smooth stellar halo, having systematically higher metallicity and lower [$\alpha$/Fe], as might be expected for more recently accreted objects \citep{font08}.
Analogous studies have also found these variations in M31 \citep{richardson08,gilbert09,bernard15}.

\subsubsection{Recovering accretion histories\index{recovering accretion histories}}
\label{sec:history}

If stars, which may now spread throughout our dark matter halo, can be connected in such a way as to reassemble their original associations with infalling satellites, then the understanding of debris evolution outlined in Sect. \ref{sec:principles} might be applied to learn about the original masses, orbits and infall times of those satellites.
Collectively, these reconstructed groups might tell us  the accretion history of our Galaxy from the stellar halo.

Several approaches have been proposed to attempt this reconstruction.
Conceptually, the simplest is to take a sample of a single type of star (i.e. with restricted absolute magnitude range) from a large-scale photometric catalogue (e.g. M-giant stars from 2MASS or MSTO and BHB stars from SDSS) and search for groups in the 3-D space of angular position and apparent magnitude \citep[e.g.][]{sharma10}.
This approach is only effective for more recent accretion events (last several billion years) since earlier events have time to phase-mix and are not apparent as separate spatial groups.
Exactly how far back in time, and the lowest luminosity of objects that might be recovered depends on the scale and depth of the survey as well as the stellar population that it is sensitive to \citep{sharma11}.

\citet{helmi01} proposed a much more powerful approach to recovering stars from early events, but also one that requires rather more data dimensions.
If the full six dimensions of phase-space can be measured for stars, then (within a given potential) their orbital properties can be calculated.
In a static potential, while they spread out over time in phase-space,  they will conserve their orbits and hence remain as a group in the space of orbital characteristics (e.g. energy and actions) indefinitely.
Several studies have analyzed prospects for {\it Gaia}\index{Gaia} in this context \citep{gomez10b,sharma11b,gomez13}.

One limitation to identifying satellite members by using observed orbital properties is defining what those properties are:
it is as yet unclear whether the Milky Way can be represented by an integrable  mass distribution in which actions can be derived.
Moreover, the potential of the Milky Way is time-evolving and this can scatter debris stars away from their original orbits.
However, stars can also ``remember where they came from'' in other ways: their chemical abundances reflect the gas cloud in which they are born.
\cite{freeman02} proposed that, given a large enough sample of high-resolution spectra of disk stars, this chemical memory could be exploited to measure the history of star formation in the disk:
while stars might be spread throughout the 6-dimensional phase-space volume occupied by the disk, those born in the same cluster would all lie at a single location in the $N$-dimensional space of chemical abundances.
These distinct  chemical abundance patterns could be used to regroup them in their original birth clusters --- an approach that  \cite{freeman02} dubbed {\it chemical tagging}\index{chemical tagging/fingerprinting}.

While stellar populations in satellite galaxies are spread out over a range of abundances (i.e. a small volume in $N$-dimensional chemical abundance space), the trends with satellite mass and assumed accretion time already seen in observations and simulations suggest that an analogous {\it chemical tagging} might work in the halo --- perhaps not to reconstruct the exact objects from which stars came, but at least to look at the numbers of satellites of a given luminosity that might have accreted onto the Milky Way at different times.
And, since the composition of a star cannot be erased by any dynamical evolution, this approach might work to recover the luminosity function of the very earliest infalling objects.
Preliminary studies of how feasible this idea is to implement in practice are only just coming to fruition \citep{lee15}.

\subsubsection{Accreted populations as a window on galaxy formation over cosmic time}
\label{sec:higherz}

Reconstructing the accretion history of our Galaxy is an exciting goal in itself, but it also opens up other possibilities.
%leads the study of satellite debris down a different direction.
If we can find the stars --- or at least identify the stellar populations --- from similar-mass, long-dead objects infalling into the Milky Way at earlier epochs, this can give us a unique window on what baryons were doing in galaxies over cosmic time.  In particular we can study baryons in the high-redshift progenitors of Milky-Way-type galaxies that may be impossible to see {\it in situ} even with the next generation of space telescopes \citep{okrochkov10}.
In fact, the stellar populations in the Milky Way's halo today were originally formed in potential wells of many different depths (from the expected mass-spectrum of infalling dark matter halos) and that formed stars for different lengths of times (dictated by the spread in accretion times for those smaller halos onto the main Milky Way halo).
Moreover, these  infalling halos may have formed in a variety of environments as later infalling objects are expected to be spread out over a larger volume at early times compared to earlier infalling objects \citep[see][]{corlies13}.
Indeed, the difference between the abundance patterns in low-metallicity stars in the halo and those observed in several ultra-faint dwarf satellites of the Milky Way can be attributed to differences in the degree to which they evolved in chemical isolation in the early Universe \citep[see][ for a discussion]{lee13}. 
Hence, the study of detailed chemical abundances of stars in the halo can tell us not just about our own Galaxy, but about the properties of stellar populations in many smaller galaxies over cosmic time.
%Several different populations.... smooth halo, substructures, satellites.....\citet{corlies,lee}

\section{Summary of status and prospects}
\label{sec:status}

Overall, 
%the last two decades of 
the discovery (over the last two decades) of debris structures encircling our Galaxy has motivated the development of a fairly sophisticated understanding of and tools for modeling debris structures. 
Putting these models and observations in a cosmological context has led to a dramatic, local confirmation of the hierarchical contribution to structure formation on small-scales: there is broad agreement between the models and data with the picture that a large fraction of our stellar halo results from the accretion of smaller systems. 
However, the wide variety of possible accretion histories means that this demonstration itself, while interesting, does not place strong constraints on cosmological models.

Two aspects of studies of debris around the Milky Way remain ripe for further exploration with near-future data.
First, 
%the possibility of recovering 
we can use stars from disrupted satellites to recover the accretion  history of our Galaxy.
The unbound structures apparent in current data sets represent the most dominant events accreted in the last several billion years.
%Near-future 
Data sets that will be available in the near future will enable the identification of debris that is either older (i.e. more fully phase-mixed) and from lower luminosity objects, both because of the large numbers of stars that will be catalogued (e.g. by {\it Gaia}) and because of the additional data dimensions that will have accurate measurements (e.g. proper motions from {\it Gaia} and detailed chemical abundances from the {\it GALAH} survey).
Most importantly, the additional data dimensions will allow the calculation of quantities that are likely to be conserved during the lifetime of the stars (e.g. orbital properties such as energy and actions as well as chemical composition), which will further enable the ``tagging'' of these stars into the groups in which they originally formed.
We can look forward to using these approaches to look much further back to the very earliest epochs of galaxy formation using stellar surveys.

While the accretion history of our Galaxy will play a key part in understanding the specific story of its formation and evolution, it is unlikely to have broader implications for the histories of other Milky-Way-sized galaxies, which are expected to have accretion histories that vary widely.
However, a by-product of the accretion history will be the luminosity function of smaller infalling galaxies, along with the identification of stellar populations associated with these different mass objects looking back to the higher redshifts at which they were accreted.
Such small galaxies will be very hard (if not impossible) to observe at higher redshift even with the next generation of telescopes.
Hence the second aspect of debris studies that should be further explored is what we can learn about the evolution of dwarf galaxies over cosmic time from our stellar halo. 
In particular, what can the stellar populations tell us about the baryonic physics of galaxy formation within small dark matter halos --- the processes of gas inflow, star formation, and feedback that remain key challenges in understanding galaxies in the Universe today?

\begin{acknowledgement}
KVJ gratefully acknowledges David Hendel's work in generating the simulations and related figures used in section 3 of this chapter.
KVJ thanks her postdocs and graduate students for invaluable discussions through
out the year (Andrea Kuepper, Allyson Sheffield, Lauren Corlies, Adrian Price-Whelan, David Hendel and Sarah Pearson).
Her work on this volume was supported in part by NSF grant AST-1312196.
\end{acknowledgement}
%
%\section*{Appendix}
%\addcontentsline{toc}{section}{Appendix}
%
%
%When placed at the end of a chapter or contribution (as opposed %to at the end of the book), the numbering of tables, figures, and equations in the appendix section continues on from that in %the main text. Hence please \textit{do not} use the \verb|appendix| command when writing an appendix at the end of your chapter or contribution. If there is only one the appendix %is designated ``Appendix'', or ``Appendix 1'', or ``Appendix 2'', etc. if there is more than one.

%\input{referenc}

%\backmatter
%\printindex
%
\end{document}